\documentclass[12pt]{article}

\usepackage{amssymb,fullpage}

\input{Qcircuit}

\newcommand{\A}{\mathcal A}

\newcommand{\B}{{\mathrm B}}
\newcommand{\bB}{{\mathbb B}}
\newcommand{\R}{{\mathbb R}}
\newcommand{\N}{{\mathbb N}}
\newcommand{\C}{{\mathbb C}}
\newcommand{\cH}{{\cal H}}
\newcommand{\cA}{{\cal A}}
\newcommand{\cP}{{\cal P}}
\newcommand{\<}{\langle}

\newcommand{\lra}{\rightarrow}

\renewcommand{\>}{\rangle}
\newtheorem{Definition}{Definition}

\newcommand{\band}[2]{
\begin{array}{|cc|}
\hline
#1 & #2 \\
\hline
\end{array}
}

\newcommand{\doubleband}[4]{
\begin{array}{|cc|}
\hline
#1 & #2 \\
#3 & #4 \\
\hline
\end{array}
}

\newcommand{\I}{\mathrm{I}}
\renewcommand{\H}{\mathrm{H}}
\renewcommand{\S}{\mathrm{S}}
\newcommand{\T}{\mathrm{T}}
\newcommand{\G}{\mathrm{G}}
\newcommand{\F}{\mathrm{F}}

\newcommand{\bI}{\mathbb{I}}
\newcommand{\bH}{\mathbb{H}}
\newcommand{\bS}{\mathbb{S}}
\newcommand{\bT}{\mathbb{T}}
\newcommand{\bG}{\mathbb{G}}
\newcommand{\bF}{\mathbb{F}}

\title{A PromiseBQP-complete String Rewriting Problem}

\author{
Dominik Janzing\thanks{School of Electrical Engineering and Computer Science, University of Central Florida, Orlando, FL 32816, USA. Electronic address: \texttt{janzing@ira.uka.de}}\, and 
Pawel Wocjan\thanks{School of Electrical Engineering and Computer Science, University of Central Florida, Orlando, FL 32816, USA. Electronic address: \texttt{wocjan@cs.ucf.edu}}}
\date{Januar 17, 2008}

\begin{document}
\maketitle

\abstract{We consider the following combinatorial problem. 
We are given three strings $s$, $t$, and $t'$ of length $L$ over some fixed finite alphabet and an integer $m$ that is polylogarithmic in $L$.
We have a symmetric relation on substrings of constant length that specifies which substrings are allowed to be replaced with each other. 
Let $\Delta (n)$ denote the difference between the numbers of possibilities to obtain $t$ from $s$ and $t'$ from $s$ after $n \in \N$ replacements. The problem is to determine the sign of $\Delta(m)$. 

As promises we have a gap condition and a growth condition.  The former states that $|\Delta (m)| \geq \epsilon\,c^m$ where $\epsilon$ is inverse polylogarithmic in $L$ and $c>0$ is a constant.  The latter is given by $\Delta (n) \leq c^n$ for all $n$. 

We show that this problem is PromiseBQP-complete, i.e., it represents the class of problems which can be solved efficiently on a quantum computer.}

\section{Introduction} 
Important computer-scientific challenges for quantum information science are to discover efficient quantum algorithms for interesting problems and to understand the fundamental capabilities and limitations of quantum computation in comparison to those of classical computation.  To approach these challenges it is important to better understand the complexity class BQP,
the class of language recognition problems which can be decided efficiently on a quantum  computer.  For technical reasons it is necessary to consider a larger class, called PromiseBQP, as BQP is not known to contain complete problems.  The difference between BQP and PromiseBQP is that in the latter only inputs from a certain subset of strings, the promise, have to be correctly classified as yes or no instances, whereas in BQP the promise is always the set of all strings (see \cite{Goldreich,DiagonalEntryToC} for more details on this difference). 

Any computational model that is universal for quantum computing immediately leads to a complete problem for PromiseBQP; namely the problem to simulate that model and to determine the output. The set of allowed input strings is then given by those for which the probability to obtain the answer {\rm YES} can be guaranteed to be either greater than some fixed value or smaller than another value.  So one can interpret the results proving that models such as adiabatic, topological, or one-way quantum computing are universal to be proving that the corresponding simulation problems are PromiseBQP-complete.  However, such purely ``quantum'' problems do not really help us to understand the difference between quantum and classical computation.  Besides we already know that a universal quantum computer could be used to simulate quantum systems.  For this reason, an important challenge for complexity theory is to construct problems that seem as classical but still characterize the power 
 of quantum computation. 

We now give a short overview of PromiseBQP-complete problems\footnote{Strictly speaking, these problems were not defined as promise problems.  However, it is not to difficult to show that it is possible to reformulate them as promise problems and to show that they are PromiseBQP-complete.}.  Ref.~\cite{KnillQuadr} characterizes the class PromiseBQP by the problem to determine the sign of  so-called quadratically signed weight enumerators.  These enumerators are polynomials whose coefficients are defined via some specific quadratic form with matrices and vectors containing the entries $0,1$.  Another PromiseBQP-complete problem is formulated in the context of knot-theory.  This problem consists in approximately evaluating the Jones polynomial of the plat closure of braids at certain roots of unity \cite{PawelYard,aharonov-2006-}.  This connection between quantum computing and knot theory is not unexpected since knot theory was successfully applied to topological quantum field 
 theories and quantum computers were shown to be able to efficiently simulate topological field theories and vice versa \cite{Freedman,Jones}.  Extending the ideas of \cite{aharonov-2006-} it was shown in \cite{AharonovTutte} that the problem to approximately evaluate the Tutte polynomial at certain points of the Tutte plane is also PromiseBQP-complete.

In \cite{DiagonalEntryToC} we presented the following  simple PromiseBQP-complete matrix problem: given a real symmetric sparse matrix $A$ determine the sign of an entry  $(A^m)_{ij}$.  The promise is that absolute value is of the entry $(A^m)_{ij}$ at least $\epsilon b^m$. Here $m$ is polynomial and $\epsilon$ is inverse polynomial in the logarithm of the size of $A$ and $b$ is an a priori given upper bound on the operator norm of $A$.  Moreover, we showed that the estimation of entries remains PromiseBQP-hard when restricted to matrices with $-1$, $0$, and $1$ as entries.  

There are two main reasons why we want to improve upon our previous results.  First, we seek to give the problem a more combinatorial flavor by considering only $(0,1)$-matrices, i.e., adjacency matrices of graphs. In \cite{DiagonalEntryToC} we required negative entries in order to be able to simulate {\it interference} which is likely to be essential for quantum computing.  As it will become clear later, we can avoid all negative entries and ``simulate'' them by {\it comparing} entries of powers of $(0,1)$-matrices.  Second, we want  to describe a setting where the sparseness condition occurs
in a natural way. Recall that we refer to a definition of sparseness which implies that there is an efficiently computable function specifying the positions and the values of the non-zero entries. In the present paper such an efficiently computable function is directly
given by rules determining the edges of the graph.  The vertices of the graph are strings and two vertices are defined to be adjacent if they can be obtained from each other by a substitution of substrings; which substrings can be replaced is specified by a given relation.  

To place our result into the context of known results on the relation between BQP and classical complexity classes we should mention that
BQP is contained in the classical counting complexity class AWPP
\cite{FR98}. Here we relax 
the problem of counting the number of certain walks in a graph to an {\it approximate} estimation of the difference between two combinatorial numbers of this kind.  The crucial point was to specify the demanded accuracy so that the corresponding problem is not only PromiseBQP-hard but also {\it in} PromiseBQP. 

The paper is organized as follows. In section~\ref{R} we define the problem formally and in section~\ref{InBQP}
we show that it can be solved efficiently on a quantum computer. This section resembles the corresponding section 
in \cite{DiagonalEntryToC}.  Nevertheless, the fact that our string rewriting problem is in PromiseBQP does not strictly follow from any of our earlier results since our definition involves a promise that generalizes the upper bound $b$ on the norm 
of the matrix $A$ in \cite{DiagonalEntryToC}.  For this reason, String Rewriting is not really an instance of our problem of estimating entries of sparse matrices.  It appears as an instance of the latter only if one neglects the promises. 
For this reason, we had to rework the entire argument. 

The main part of the paper is Section~\ref{BQPhard} where we show that String Rewriting is PromiseBQP-hard.  The idea is strongly motivated by physical intuition.  We first encode a quantum circuit solving a PromiseBQP-complete problem into
a one-dimensional translational invariant interaction with finite  interaction range.  This interaction ``Hamiltonian''  $\tilde{H}$ describes a physical system whose autonomous time evolution runs the quantum computation.  We construct the Hamiltonian in such a way that an entry of  $\tilde{H}^m$ for  appropriate $m$ determines the solution of the computation. 
Various Hamiltonians of this kind have been described in the literature in a different context.  
 The special feature of our construction of the Hamiltonian $\tilde{H}$ distinguishing it from the known ones is that 
it acts (up to a scaling factor) like a matrix with entries $0,1,-1$.
This makes it possible to define a $(0,1)$ matrix $A$ in such a way that we can express the entries of $\tilde{H}^m$ by differences of entries of $A^m$.  The fact that $\tilde{H}$ is translation invariant and has only finite interaction range is important to represent the constructed adjacency matrix by a string rewriting problem.

\section{Definition of the string rewriting problem}

\label{R}
Simply speaking, we consider the problem to estimate the difference between two combinatorial quantities: Consider three strings $s$, $t$, and $t'$ of equal length and a set of allowed modifications. These modifications correspond to replacements of substrings according to a given relation on the set of substrings. The problem is to decide whether there are more possibilities to obtain $t$ from $s$ or to obtain $t'$ from $s$ using exactly $m$ substitutions.
Let $\Delta_{s,t,t'}(m)$ denote the difference between these numbers of possibilities.  The problem is to determine its sign.  However, we have to carefully describe the given promise.  On the one hand, we have a lower bound on the absolute value of $\Delta_{s,t,t'}(m)$ for the specific $m$.  This makes it possible to determine the sign of $\Delta_{s,t,t'}(m)$ even though only an {\it approximation} of the actual value is known.  On the other hand, we have an upper bound on the growth of $\Delta_{s,t,t'}(n)$ for all $n$. Later, it will become evident that our estimation procedure for $\Delta_{s,t,t'}(m)$ is quite sensitive to this bound. 

We define our PromiseBQP-complete decision problem as follows.
\begin{Definition}[String rewriting]\label{def:stringRewriting}${}$\\
We are given three strings $s$, $t$ and $t'$ of length $L$ over some finite alphabet $\A$ and a positive integer $m=poly(L)$ as input.
Moreover, we are given a relation $\sim$ on substrings of length at most $k$ where $k$ is some constant\footnote{To show that string rewriting is PromiseBQP-hard it suffices to consider a relation with the following property.  If $u\sim v$ then $|u|=|v|$, i.e., both substrings have the same length.  It can be shown that if the relations identify substring of different lengths, then the corresponding string rewriting problem is still in PromiseBQP.}.  Replacing substrings $u,v$ with $u\sim v$  naturally gives rise to possible conversions of strings in $\A^L$. Let $A$ be the adjacency matrix of the graph whose vertices are strings in $\A^L$ and whose edges are given by the possible conversions.  For $n\in\N$ let
\[
\Delta_{s,t,t'}(n):=A^n_{s,t}-A^n_{s,t'}
\]
be the difference between the number of possibilities to obtain $t$ from $s$ and those to obtain $t'$ from $s$ after exactly $n$ replacements. 
 
We are given the promises 
\begin{equation}\label{Growth}
\Delta_{s,t,t'}(n)\leq c^n \quad \mbox{for all } n \in \N
\end{equation}
and $|\Delta_{s,t,t'}(m)|\geq \epsilon c^m$ where $c>0$ is a constant and $\epsilon\in [0,1]$ with $\epsilon=1/poly(L)$.  The problem is to decide if either $\Delta_{s,t,t'}(m)> 0$ or $\Delta_{s,t,t'}(m)<0$.
\end{Definition}

It should be mentioned that the relation on the substrings is not necessarily an {\it equivalence} relation. Our construction of the PromiseBQP-hard instance in section~\ref{BQPhard} will, for instance, be based on a non-transitive relation. 

Our formulation of the problem deserves some explanations.  
In a regular graph such a difference between the number of walks can be directly 
interpreted as probability differences in a discrete time random walk: If $d$ is the degree,  $(A^m)_{s,t}/d^m$ is the probability to arrive at $t$
given that $s$ was the position $m$ steps earlier. 
Having this interpretation in mind, it is certainly natural to ask whether random replacements are more likely to lead to string $t'$ or to string $t$ after $m$ steps.
In order to obtain a decision problem, we need a promise on the absolute value of the difference so that even an inaccurate estimation can distinguish between both cases.  This is just the usual way how promises are used to cast estimation problems as decision problems \cite{Goldreich}.  To understand the role of the growth condition in ineq.~(\ref{Growth}) is more difficult.  Roughly speaking, it implies some statements on the spectrum of $A$ making it possible to adjust the algorithm to the smallest scale on which $\Delta_{s,t,t'}(m)$ can be estimated.

\section{The string rewriting problem is in PromiseBQP}

\label{InBQP}
The main contribution of this paper is to show that our string rewriting problem is PromiseBQP-hard.  The proof that it is in PromiseBQP is based on quantum phase estimation, following closely the proofs in \cite{DiagonalEntryToC,NumberOfWalks} .  However, we emphasize that it cannot be directly reduced to the problems in these references since the promises of the problems are not compatible. Nevertheless, the idea of the the proof is quite similar but the quantum algorithm presented here is more general. 

Let us briefly rephrase the definition of PromiseBQP.  The characteristic feature of promise problems is that a set 
$\Pi \subseteq \{0,1\}^*$ specifies the allowed input strings. 
\begin{Definition}[The complexity class PromiseBQP]\label{promiseBQP}${}$\\
PromiseBQP is the class of promise problems $(\Pi_{{\rm YES}},\Pi_{{\rm NO}})$ that can be solved by a uniform\footnote{By ``uniform'' we mean that there exists a
polynomial time classical algorithm that generates a sequence of a polynomial number  of  quantum gates for every desired input length.} family of quantum circuits. More precisely, it is required that there is a uniform family of quantum circuits $Y^{(r)}$ acting on ${\rm poly}(r)$ qubits that decide if a string $\bf{x}$ of length $r$ is a YES-instance or NO-instance in the following sense.  The application of $Y^{(r)}$ to the computational basis state $|{\bf x},{\bf 0}\rangle$ produces the state 

\begin{equation}\label{Schalt}
Y^{(r)} |{\bf x},{\bf 0}\> = 
\alpha_{{\bf x},0} |0\> \otimes |\psi_{{\bf x},0}\> + 
\alpha_{{\bf x},1} |1\> \otimes |\psi_{{\bf x},1}\>
\end{equation}
such that
\begin{enumerate}
\item for every ${\bf x}\in\Pi_{{\rm YES}}$ it holds that $|\alpha_{{\bf x},1}|^2 \geq 2/3$ and
\item for every ${\bf x}\in\Pi_{{\rm NO}}$ it holds that $|\alpha_{{\bf x},1}|^2 \leq 1/3$.
\end{enumerate}
Any output is acceptable if ${\bf x} \not\in \Pi=\Pi_{{\rm YES}}\cup \Pi_{{\rm NO}}$.  
Equation~(\ref{Schalt}) has to be read as follows. The input string
${\bf x}$ determines the first $r$ bits. Furthermore,  $k$
additional ancilla bits are initialized to $0$. After $Y^{(r)}$ has been
applied we interpret the first qubit as the relevant output and the
remaining $r+k-1$
output values are irrelevant. The size of the
ancilla register is polynomial in $r$.
\end{Definition}

BQP is the class of problems in PromiseBQP for which
the promise is trivial, i.e., $\Pi$ is the set of all strings.

To show that our string rewriting problem is in PromiseBQP we first introduce some notation and terminology. 
Let 
\[
A=\sum_j \lambda_j Q_j
\]
be the spectral decomposition of the adjacency matrix $A$ and $|\varphi\>$ be an arbitrary quantum state.  The spectral measure induced by $A$ and $|\varphi\>$ is a probability distribution on the spectrum of $A$ such that the eigenvalue $\lambda_j$ occurs with probability 
$\langle \varphi |Q_j|\varphi\>$.  We refer to any quantum process allowing us to sample from the spectral measure as {\it measuring the observable $A$ in the state} $|\varphi\>$.  This corresponds to {\it von Neumann measurements} of quantum observables.  
To solve our problem on a quantum computer we will identify strings $s$ with basis vectors $|s\rangle$ of  a quantum register. This can be done in a straightforward way by representing each symbol by a basis state on $\lceil log_2(|\A|) \rceil$ qubits. 

The first basic observation making it possible to solve the string rewriting problem is that 
for any self-adjoint matrix $A$  measuring 
 $A^m$ can be reduced to measuring $A$ since we can compute
$\lambda^m$ from the outcome $\lambda$. Due to standard spectral theory we have
\[
\<\varphi|A^m|\varphi\> = \sum_j \lambda_j^m\langle \varphi|Q_j|\varphi\>\,,
\]
i.e., the left hand side is the $m$th statistical moment of the spectral measure induced by $A$. 
We can estimate this value by repeatedly applying $A$-measurements to the state $|\varphi\rangle$ and computing the average
of the $m$th power of the obtained outcomes. 
The second observation  is that $\Delta_{s,t,t'}(m)$ is equal to
\begin{equation}\label{linComb}
\Delta_{s,t,t'}(m) = A^m_{st}-A^m_{st'} =\sqrt{2}\<s|A^m|\varphi \>=\sqrt{2}\Big(\langle \psi^+| A^m|\psi^+\rangle -\langle \psi^-|A^m|\psi^-\rangle\Big)\,, 
\end{equation}
where we have defined 
\[
|\varphi\rangle :=\frac{1}{\sqrt{2}}(|t\rangle -|t'\rangle)
\]
and
\[
|\psi^\pm\rangle:=\frac{1}{\sqrt{2}}(|s\rangle \pm |\varphi\rangle )\,.
\]
The second equality in eq.~(\ref{linComb}) is obvious, the third one is the polarization identity for real quadratic forms. Eq.~(\ref{linComb}) shows that we can write the difference to be estimated as a difference between the $m$th statistical moments of the  probability distributions induced by measurements applied to the states $|\psi^{\pm}\>$.

The algorithm consists of the following three steps.  Later, we will modify the third step to make the scheme less sensitive to errors.

\begin{enumerate}
\item Given an upper bound $b$ on $\|A\|$, construct a quantum circuit implementing the unitary operator $U:=\exp(-iA/b)$ approximately. This is known to be possible for all sparse matrices $A$ \cite{ATS,ChildsDiss,BACS:06}.  

\item To realize phase estimation for $U$ proceed as follows.  Replace the gates of $U$ with corresponding controlled gates to obtain the controlled-$U$ gate.  Repeat this transformation $j$ times to obtain the controlled-$U^j$ unitary matrices.   The outputs of phase estimation are the eigenvalues of $U$ uniquely identifying those of $A$.  Hence we have an approximate implementation of the von-Neumann measurement of the observable $A$. 

\item Apply $A$-measurements to the states $|\psi^\pm \>$.  By raising each outcome to the $m$th power,
we estimate the corresponding expected values of $A^m$ in these states by computing the average over repeated sampling. 
To see that an efficient preparation procedure for $|\psi^\pm\rangle$ exist we observe that we could, for instance, generate
superpositions between the first three basis states and apply any unitary operation that maps these states to $|s\rangle$, $|t\rangle$, and $|t'\rangle$. 
\label{Schrittm}

\end{enumerate}

There are two delicate points deserving our attention.  First, we can only implement an {\it approximate} measurement of the observable $A$.  By taking the $m$-th power of the outcomes the inaccuracies could be amplified up to the scale given by the $m$th power of the largest measurement outcomes that are possible.  Second, the statistical error caused by the fact that we estimate expected values after a polynomial number of samples can only be bounded from above if we have prior knowledge on the probability of large eigenvalues.  
For these reasons we modify the above scheme in order to make it less sensitive to the occurrence of large measurement outcomes and to small fluctuations of these values. 
We rewrite eq.~(\ref{linComb}) as
\begin{equation}\label{DiffSpec}
\Delta_{s,t,t'}(m) =\sqrt{2} \sum_j (p^+_j-p^-_j) \lambda_i^m\,.
\end{equation}
where $p^\pm_j$  denote the probabilities of the eigenvalue $\lambda_j$ in the spectral measure induced by 
$|\psi^+\rangle$ and $|\psi^-\rangle$, 
respectively.  Note that the growth condition in eq.~(\ref{Growth}) ensures that $p^+_j-p^-_j=0$ for all $j$ with $\lambda_j>c$.  If we could sample from the spectral measure exactly, we could just drop all outcomes $\lambda$ with $\lambda>c$.  But we can only achieve that
the outcomes are  {\it close} to the true eigenvalues {\it with high probability}.

Let us define a function $f$ by
\[
f:\R \rightarrow [-c^m,c^m]\,, \quad
f(x):= \left\{ 
\begin{array}{rcccr} 
x^m  & \hbox{for} &  |x| &  \leq & c   \\  
c^m  & \hbox{for} &   x  &  >    & c   \\ 
-c^m & \hbox{for} &   x  &  >    & -c 
\end{array} 
\right.
\]
We can replace $\lambda_j^m$ in eq.~(\ref{DiffSpec}) with $f(\lambda_i)$.  Hence we have rewritten $\Delta_{s,t,t'}(m)$ as a linear combination of expectations of $f(A)$ instead of $A^m$.  Estimating the expected value of $f(A)$ by measuring $A$ is less sensitive to errors than estimating $A^m$.  This is because $f$ is Lipschitz continuous with constant $m\, c^{m-1}$.  
The following argument shows that this implies sufficient accuracy for estimating the expected values of $f(A)$.
Assume for the moment that we apply the $A$-measurement to an eigenvector of $A$ with eigenvalue $\lambda_j$. We showed in \cite{DiagonalEntryToC,NumberOfWalks} that there is an efficient method that guarantees the following precision. With probability $1-\theta$ the outcome $\lambda$ is $\eta$-close to $\lambda_j$, where $\eta$ and $\theta$ 
are inverse polynomial in the description length of the matrix $A$. Consequently, the probability that $\lambda^m$ deviates from $\lambda_j^m$ by more than 
$\eta\, m\, c^{m-1}$ 
is at most $\theta$. This provides the following upper bound on the discrepancy between the expected value in our imperfect measurement and the expected value of a perfect measurement applied to the eigenvector:
\begin{equation}\label{Acc}
\eta \, m \,c^{m-1} + 2 \,c^m \theta\,.
\end{equation}
 By linearity (see the analogue arguments in \cite{DiagonalEntryToC,NumberOfWalks}), the same estimation also holds for general states. By choosing $\theta$ and $\eta$ appropriately the error is at most $\epsilon c^m/2$ for every inverse polynomial $\epsilon$. Using Hoeffding's inequality in straightforward analogy to the estimations of subsection 5.3 in \cite{DiagonalEntryToC} one sees that we can also make the error caused by finite sampling smaller than $\epsilon c^m/2$ using a polynomial number of runs. Hence we obtain an estimation of  $\Delta_{s,t,t'}(m)$ up to an error of $\epsilon c^m$ with any desired probability that is inverse polynomially close to $1$. 

It is straightforward but technical to construct a quantum circuit including the computation of the averages after repeated sampling such that it solves String Rewriting in the sense of Definition~\ref{promiseBQP}.

\section{String rewriting is PromiseBQP-hard}\label{BQPhard}
To prove that our problem is PromiseBQP-hard we have to show how to encode a quantum circuit into a graph that is derived from string rewriting rules in the sense of Definition~\ref{def:stringRewriting}. The proof is divided into four parts.  First, we encode it into a translationally invariant $3$-local Hamiltonian $H$ acting on a qudit chain. Second, we establish that a certain entry of $H^m$ (an appropriate power of this Hamiltonian) indicates the solution of the quantum circuit.  Third, we convert this Hamiltonian $H$ into 
an operator $\hat{H}$ that acts like an adjacency matrix $A$ on the relevant subspace 
such that the difference of certain two entries of $A^m$ indicates the solution.  Fourth, we show how to choose the exponent $m$. 

\subsection{The Hamiltonian $H$} 
Let $Y^{(r)}$ be a quantum circuit that solves a PromiseBQP-complete problem as in Definition~\ref{promiseBQP}. Define $U$ by
\[
U:=Y^{(r)} \sigma_z {Y^{(r)}}^\dagger\,,
\] 
where $\sigma_z$ is the Pauli matrix acting on the output qubit. Elementary calculation shows \cite{DiagonalEntryToC} that 
\begin{equation}\label{overlap}
\langle {\bf x},{\bf 0}|U|{\bf x},{\bf 0}\rangle=p_0-p_1  \,,
\end{equation}
where $p_j:=|\alpha_{{\bf x},j}|^2$ is the probability of finding the output qubit in the state $|j\>$ for $j=0,1$. We now construct a Hamiltonian $H$ such that $\<\alpha |H^m|\omega\>$ is equal to $p_0-p_1$ in eq.~(\ref{overlap}) times some constant. Here $\alpha$ and $\omega$ define strings that represent the initial and the final states of our computation, respectively.  Our goal to 
express certain entries of
 the powers of the Hamiltonian $H$ as entries of the  powers of an adjacency matrix imposes very special requirements on its form. This prevents us from using the construction in \cite{Vollbrecht} since it is essential for our purposes to work only with Toffoli and Hadamard gates. Note that similar constructions have been described in the literature in a different context. The constructions in \cite{Benioff,Feynman:85,Margolus:90} showed that Hamiltonian time evolutions can in principle perform computations (which had also been an issue of thermodynamics of computation at that time) and those in \cite{Ergodic,ErgodicQutrits} that even relatively simple Hamiltonians can perform universal quantum computing. Moreover, there are construction of Hamiltonians whose low-energy states encode QMA-complete problems \cite{KitaevShen,Kempe2local,Oliveira,GottesmanQMA}.

Neither of the above constructions satisfies all the conditions required for our proof: we need a $1$-dimensional translationally invariant Hamiltonian acting on a qudit chain with finite range interactions whose autonomous time evolution makes it possible to realize universal quantum computing and is spectrally isomorphic to a quantum walk on a line. The Hamiltonian $H$ that we now construct will, at first, follow 
our construction in  \cite{CMP} (which was used to prove that a certain quantum measurement problem is PromiseBQP-hard)
but will be modified later. It acts on a qudit chain of length $L$, i.e., on the Hilbert space $\cH$ given by
\[
\cH:=\cH_c^{\otimes L}
\]
where $\cH_c$ is the Hilbert space of a single cell. Each cell consists of a data and a program cell. Their Hilbert spaces are given by $\cH_d=\C^{4}$ and $\cH_p=\C^{14}$, respectively. The basis vectors of $\cH_d$ are identified with the symbols
\begin{equation}\label{data}
\{0,1\} \cup \{\,\|,\bullet\}
\end{equation}
and those of $\cH_p$ with
\begin{equation}\label{program}
\{\I,\H,\S,\T,\B\} \cup \{\bI,\bH,\bS,\bT,\bB\} \cup \{\square,\blacksquare, \Diamond, \#\}\,.
\end{equation}
We construct an operator $V$ acting on $\cH_c^{\otimes 3}$ and use it to define our translationally invariant $3$-local Hamiltonian $H$. 
The form of $H$ is 
\[
H= F+F^\dagger=\sum_j (V_j + V^\dagger_j) \,,
\]
where $V_j$ denotes the embedding of $V$ such that $V_j$ acts on cells $j-1$, $j$ and $j+1$ of $\cH_c^{\otimes L}$. Accordingly, we address the three cells on which $V$ acts by $-1$, $0$, and $1$. 

In analogy to the constructions mentioned above, the idea is that the {\em forward time operator} $F$ simulates the computation in the sense that $F^m|\alpha\>$ represents the $m$th step of the computation. The program band contains the program code specifying the sequence of gates to be applied to the data register, a certain region of the data band. A block of $n$ program cells corresponds to one time step (layer) of a parallelized circuit. Such a time step can consist of an application of several gates acting on disjoint qubit sets. To simulate one time step of the circuit, the code has to be first translated to the left by $n$ cells so that the commands of that time step are exactly above the data register and then the commands have to be executed.  All these operations have to be performed by the local action of the copies of $V$. 

$V$ is a concatenation of two operators
\[
V = C \, X\,.
\]  
The {\em execution operator} $X$ implements logical transformations on the data band that are controlled by the state of the program band. These logical transformations include the implementation of proper quantum gates. The {\em transition operator} $C$ implements transitions between configurations of the program band. These transitions realize the propagation of the program. Before we formally define the operators $X$ and $C$ we first describe 
the intuition behind how they are used to implement the computation.

\begin{enumerate}

\item Without loss of generality we assume that $U$ consists of Toffoli gates acting on triples of adjacent qubits, swap gates acting on pairs of adjacent qubits, and Hadamard gates.  Moreover, we assume that the first two qubits are the control qubits of the Toffoli gate and the third qubit is its target qubit. This is because the Toffoli and the Hadamard gates form a universal set of gates provided that the Toffoli gate can act on triples of arbitrary qubits \cite{aharonov-2003} and non-adjacent qubits can be brought together by applying a sequence of swap gates. 

Assume we want to use $F$ to to simulate the computational process defined by the following quantum circuit $U$

\medskip
\begin{center}
\hspace{2cm}
\Qcircuit @C=1em @R=.7em {
\lstick{|b_1\>} & \ctrl{2} & \gate{H}         & \qw \\
\lstick{|b_2\>} & \ctrl{1} & \multigate{1}{S} & \qw \\
\lstick{|b_3\>} & \targ    & \ghost{S}        & \qw \\
\lstick{|b_4\>} & \gate{H} & \qw              & \qw 
}
\end{center}
\medskip

\noindent and also require that the simulation immediately stops after the swap gate has been applied.  In this case, we would initialize the program and the data bands as follows:
\[
\begin{array}{ccccccccccccccccccc}
\multicolumn{7}{c}{} & \multicolumn{3}{c}{\mbox{(first step)}} &  & &\multicolumn{4}{c}{\mbox{(second step)}} \\
\#      & \#   & \#      & \#      & \#      & \#      & \B & \blacksquare & \I  & \T  & \H  & \I & \H      & \I      & \S      & \B      & \I  & \#      & \hbox{ (program) }\\
\bullet & \|   & \bullet & \bullet & \bullet & \bullet & \| & b_1          & b_2 & b_3 & b_4 & \| & \bullet & \bullet & \bullet & \bullet & \|  & \bullet & \hbox{ (data) }
\end{array}\,
\]

The program code consists of a sequence of several gate symbols $\I$, $\S$, $\T$, $\H$, one execution symbol $\blacksquare$, and two boundary symbols $\B$. 
The gate symbols  trigger the implementation of the corresponding gates.
The blank symbol $\#$ fills the empty program cells. The initial logical state $|{\bf  x},{\bf 0}\rangle$ is  written into the data cells marked by $b_1,\dots,b_4$ using the symbols  $0,1$. 
The symbols $\bullet$ and $\|$ to the left and to the right of the data register are used to control the propagation and execution of the program code. The formatting  symbol  $\|$ divides the data band into blocks. This makes it possible  to ensure 
that the program is only executed after the program  has  been propagated by another $n$ sites (one block) such  that the next 
time step (i.e., a block of $n$ commands) of the simulated circuit is implemented. The symbol $\bullet$ marks the remaining unused data cells.

Initially, the first block of the program code is aligned with the data block 
in the sense that the gate symbols corresponding to the first time step of  $U$ 
are located above the data cells they act on. 
 
\item The execution symbol $\blacksquare$ is propagated cell by cell to the end of the program code. In each step, the symbol $\blacksquare$ and the gate symbol $\G$ to the right of it are swapped and the gate corresponding to $\G$ is applied to the relevant data cells. Note that in the initial configuration the execution symbol $\blacksquare$ is at the right of the left end boundary symbol $\B$. This is because the only purpose of the left copy of $\B$ is to ensure that a ``computation in backward direction'' (i.e., the application of $F^\dagger$) annihilates the initial state. 

\item Once the execution symbol $\blacksquare$ has passed the end of the program code it is converted to the blank symbol $\#$ via the intermediate symbol $\Diamond$ and a signal is sent left toward the begin of the program code. This signal indicates that the execution of the first time step of the program has been completed.  It propagates by  converting each gate symbol $\G$ to $\bG$ cell by cell. Once the marked gate symbol $\bG$ is at the begin of the program code it triggers the conversion of $\#$ into the hole symbol $\square$ via the intermediate 
symbol $\Diamond$.

\item The hole symbol $\square$ propagates cell by cell to the end of program where it is converted to $\#$ and triggers the creation of the left propagating marker $\bG$ via the intermediate turn-around symbol $\Diamond$.  The arrival of this marked gate symbol at the begin of the program code triggers the conversion of the next copy of $\#$ into $\square$. This procedure is repeated until the begin of the program code is aligned with the next copy of the format symbol $\|$ on the data band.  In this situation, the marked gate symbol $\bG$ triggers the creation of $\blacksquare$ instead of $\square$ and the whole cycle starting in step $2$ is repeated until the execution symbol $\blacksquare$ and the boundary symbol $\B$ meet above the data register, which leads to the termination of the program. 
\end{enumerate}

To define the transition operator $C$, we need to introduce the following transition rules:
\medskip
\[
\begin{array}{ccccccccccc}
1 & \mbox{a)} & \band{\square}{\G}  & \lra & \band{\G}{\square}   & & & \mbox{b)} \quad &\,\,\band{\blacksquare}{\G} & \lra & \band{\G}{\blacksquare} \\ \\
2 & \mbox{a)} & \doubleband{\square}{\#}{{*}}{}& \lra & \doubleband{\Diamond}{\#}{{*}}{} & & & \mbox{b)} \quad &\,\,\doubleband{\blacksquare}{\#}{\|}{} & \lra & \doubleband{\Diamond}{\#}{\|}{} \\ \\
3 &           & \band{\G}{\Diamond} & \lra & \band{\bG}{\#} \\ \\
4 &           & \band{\F}{\bG}      & \lra & \band{\bF}{\G} \\ \\
5 &           & \band{\#}{\bG}      & \lra & \band{\Diamond}{\G} \\ \\

6 & \mbox{a)} & \doubleband{\#}{\Diamond}{}{{*}}  & \lra & \doubleband{\#}{\square}{}{{*}} && & \mbox{b)} \quad &\,\,\doubleband{\#}{\Diamond}{}{\|} & \lra & \doubleband{\#}{\blacksquare}{}{\|} 
\end{array}
\]
We use $*$ to denote any of the symbols $\bullet$, $0$, and $1$. The symbol represented by $*$ is left unchanged by the corresponding transition rule. In the transition rules $2$ and $6$ the lower right and left corners are left empty to indicate that the symbol at that place is not important for the transition rule and that it is left unchanged.

Transitions 1 a) and b) implement the rightward propagation of the symbols $\square$ and $\blacksquare$, respectively. Transitions 2 a) and b) take place when $\square$ and $\blacksquare$ have passed the end of the program code, respectively.  They create the turn-around symbol $\Diamond$. Transition 3 creates a marked gate symbol that initiates the leftward moving signal.  Transition 4 implements the propagation of this signal.  Once this signal has arrived at the begin of the program code, transition 5 and transitions 6 a) and b) create the symbols $\square$ and $\blacksquare$ via the creation of the intermediate turn-around symbol $\Diamond$, respectively. The execution symbol $\blacksquare$ is created only if the turn-around symbol $\Diamond$ is exactly above the formatting symbol $\|$, which happens only if the blocks of the program code are aligned with the data register.  Otherwise, the hole symbol $\square$ is created.

The transition operator $C$ acts on $\cH_c^{\otimes 3}$, implemens the above transitions between configurations, and annihilates all the configurations that do not appear to the left of the arrow symbol in any of the above rules.  $C$ is embedded such that the left side of each box refers to cell $0$ and the right side to cell $1$. 

Before we define the execution operator $X$, we specify the conditions for implementing the Toffoli, Hadamard, swap gates, and ``boundary'' gates:
\begin{itemize}
\item[-] program cell $0$ contains the execution signal $\blacksquare$
\item[-] program cell $1$ contains the corresponding gate symbol $\G$
\item[-] neither $\bullet$ nor $\|$ is contained in data cells $-1$, $0$, and $1$ in the case of the Toffoli gate, in data cells $0$ and $1$ in the case of the swap gate, and in data cell $1$ in the case of the Hadamard and boundary gates.
\end{itemize}

Define $P_\G$ to be the projectors onto the subspaces of $\cH_c^{\otimes 3}$ spanned by the configurations for which the above three conditions are fulfilled for $\G\in \{\S, \T, \H, \B\}$ and
\[
P_1 := {\bf 1} - P_\S - P_\T - P_\H - P_\B\,.
\] 
Let $U_\G$ be the unitary operators that implement the gate corresponding to $\G$ on the relevant data cells for $\G\in\{\H,\S,\T\}$. More precisely, they implement the gates on the subspaces spanned by the logical symbols $0$ and $1$ in the relevant data cells and act trivially on the orthogonal complement of this subspace and on the state space of the cells of the program band. 

The execution operator $X$ is now defined by 
\begin{equation}\label{Vdef}
X := U_\T P_\T + U_\S P_\S + U_\H P_\H + P_1\,.
\end{equation}

Note that expression~(\ref{Vdef}) does not contain an operation that is executed in the subspace corresponding to $P_\B$. The operator $X$ is constructed such that it leads to annihilation when the execution $\blacksquare$ and the boundary $\B$ symbols meet above the data register for the first time.

\subsection{Encoding of solutions into entries of powers of the Hamiltonian}\label{subsec:encodingPowers}

To show that certain entries of $H^m$ encode the solution of the PromiseBQP-complete problem for appropriately chosen $m$, we denote the initial state of the program and data bands by $|\alpha\>$ and consider the vectors $F^i|\alpha\>$ that represent the different steps of the computation for $i=0,\dots,\ell-1$.  Here $\ell-1$ is the number of steps it takes till the execution symbol $\blacksquare$ and the boundary gate symbol $\B$ meet above the data register for the first time, i.e., the very next step would annihilate the state. Observe that the vectors $F^i|\alpha\>$ are mutually orthogonal and have length $1$ for $i=1,\ldots,\ell-1$.  This follows from the fact that only one transition rule can be applied to the configurations of the program band in the vectors $F^i|\alpha\>$ for $i=1,\ldots,\ell-2$ and the configurations are all different for $i=1,\ldots,\ell-1$.  

Define 
\[
|\omega\>:=|{\bf x},{\bf 0}\> \otimes |\chi\>\,,
\]
where the left component $|{\bf x},{\bf 0}\>$ is the initial state of the data register and the right component $|\chi\>$ corresponds to the final basis state of the remaining part of the data band and the whole program band in the $(\ell-1)$th step.  Note that $|\omega\>$ would be equal to $F^{\ell-1}|\alpha\>$ if the operator $V$ did not contain the execution operator $X$, i.e., if we only had $V=C$.  We now obtain 
\[
\<\alpha|F^{\ell-1}|\omega\> = \<{\bf x},{\bf 0}|U|{\bf x},{\bf 0}\> = p_0-p_1\,.
\]

The following identities hold: (1) $F^\dagger F^i|\alpha\> = F^{i-1}|\alpha\>$  for $i=1,\dots,\ell-1$ and (2) $F^\dagger|\alpha\>=0$. 
To see that (1) holds, examine every transition rule in the backward direction and check that only one rule is applicable for each configuration in $F^i|\alpha\>$ for $i=1,\dots,\ell-1$. Statement (2) holds because a backward step would first swap the positions of $\blacksquare$ and $\B$ in the initial state $|\alpha\>$ (performed by $T^\dagger$) and then annihilate the state (performed by $X^\dagger$). This is because the projectors $P_\T$, $P_\H$, $P_\S$, and $P_1$ annihilate a configuration containing the symbol pair $\blacksquare\B$ if the boundary symbol $\B$ is above a cell of the data register. As already mentioned, we have $F^{\ell}|\alpha\>=0$. These properties imply that the Hamiltonian $H=F + F^\dagger$, restricted to the span of $F^i|\alpha\>$ for $i=0,\dots,\ell-1$, is unitarily equivalent to
\[
\cP:=\sum_{j=0}^{\ell-1} |j\>\<j+1| +|j+1\>\<j|\,,
\]
i.e., the adjacency matrix of the line graph acting on $\C^{\ell}$. The state $|\omega\>$ consists of a component in the span of the vector $F^{\ell-1}|\alpha\rangle$ and a component perpendicular to the whole orbit $F^i|\alpha\>$. Its overlap with $F^{\ell-1}|\alpha\>$ (corresponding to the last basis vector in $\C^\ell$) is given by $p_0-p_1$.  This finally proves that  
\[
\<\alpha|H^m|\omega\> = (p_0 - p_1)\, (\cP^m)_{0,\ell-1}\,.
\]
We choose $m$ to be odd if $\ell$ is even and vice versa so that $(\cP^m)_{0,\ell-1}$ is strictly greater than $0$ for $m\geq \ell-1$.  Therefore, the sign of the entry $\<\alpha|H^m|\omega\>$ determines the solution of the PromiseBQP-complete problem by indicating whether $p_0-p_1$ is positive or negative.


\subsection{Derivation of the string rewriting rules}\label{Adj}

The operator $V=CX$ in eq.~(\ref{Vdef}) has only the values $1,0,\pm 1/\sqrt{2}$ as entries since $X$ has these values as entries and $T$ has only $0$ and $1$ as entries.  The former follows from the fact that the projectors in eq.~(\ref{Vdef}) act on mutually orthogonal subspaces and the unitary operators have $1,0,\pm 1/\sqrt{2}$ as entries.  The latter is due to the fact that $C$ describes transitions between basis states.  Recall that the values $\pm 1/\sqrt{2}$ stem from the Hadamard gate and the value $1$ from transitions that permute basis states. We derive the string rewriting rules by first converting $V$ to $\tilde{V}$ and then $\tilde{V}$ to $\hat{V}$
in a way that  the  corresponding Hamiltonians are still able to compute the solution.

  The operator $\tilde{V}$ contains $0,\pm 1/\sqrt{2}$ as entries and $\hat{V}$ only $0,1$. We interpret the basis states on which $\hat{V}$ operates as substrings; an entry $1$ in $\hat{V}$ then indicates that the corresponding substrings can be exchanged.  

To obtain $\tilde{V}$, we substitute the value $1$ in $V$ by matrices having only $\pm 1/\sqrt{2}$ as entries.  
This can be done by introducing an auxiliary band $\cH_a^{\otimes L}$ where $\cH_a:=\C^2$ and performing a ``dummy'' Hadamard gate accompanying each step that does not involve the execution of the Hadamard gate on the computational subspace of one of the cells of the data register. However, this additional dummy Hadamard gate is less harmless than it may seem at first glance.  Recall that in our model the output is computed from the overlap of the final state $F^{\ell-1}|\alpha\>$ and some {\it basis} state. By performing additional Hadamard gates on the auxiliary band the propagating signal would permanently trigger the creation of new superpositions 
$|\pm \>:=(|0\rangle \pm |1\rangle)/\sqrt{2}$ in the auxiliary band.  This would exponentially decrease the overlap to any basis state. 

Therefore, we must ensure that the additional dummy Hadamard gate is always applied to the {\it same} qubit.  Since the signal propagates the qubit cannot be {\it physically} the same. But it suffices to move the information with the signal. To achieve this, we distinguish between right-moving (rule 1), left-moving (rules 3, 4 and 5), and stationary transitions (rules 2 and 6). Note that these terms refer to the direction in which the signal propagates and not the direction in which the program code moves.  We denote the projections onto the subspace of $\cH_p^{\otimes 3}\otimes \cH_d^{\otimes 3} \otimes \cH_a^{\otimes 3}$ corresponding to these configurations by $Q_r$, $Q_l$, and $Q_s$, respectively. Let $S_{-1,0}$ and $S_{0,1}$ be the swap gates between the auxiliary qubits in cells $-1,0$ and those in cells $0,1$.  Let $W_\H$ denote the dummy Hadamard gate acting on the auxiliary cell $0$.  To simplify notation we assume that these swap operators, the dummy Hadamard gate and the operators $C$, $P_\G$, $U_\G$, and $P_1$ are now all embedded into $\cH_p^{\otimes 3}\otimes \cH_d^{\otimes 3} \otimes \cH_a^{\otimes 3}$.  

The modified operator $\tilde{V}$ is
\begin{equation}\label{tildeV}
\tilde{V}:=
C \,
\big(S_{0,1}\, Q_r + Q_s + S_{-1,0} Q_l\big) \,  
\big(W_\H (U_\T P_\T + U_\S P_\S + P_1) + U_\H P_\H \big)\,.
\end{equation}
The operator $\big(S_{0,1}\, Q_r + Q_s + S_{-1,0} Q_l\big)$ swaps either the auxiliary qubits in cells $-1$ and $0$ or those in cells $0$ and $1$ depending on the direction in which the signal moves.  The operator 
\[
\tilde{X}:=W_\H (U_\T P_\T + U_\S P_\S + P_1) + U_\H P_\H 
\]
applies the correct gate to data cells and the dummy Hadamard gate $W_\H$ iff no Hadamard gate is applied to the cells of the data register. 
To formally verify that $\tilde{V}$ contains only the entries $0,\pm 1/\sqrt{2}$ (as already explained in an intuitive way) 
we recall that
\[
\tilde{C} := C \, \big(S_{0,1}\, Q_r + Q_s + S_{-1,0} Q_l\big)
\] 
is a permutation  matrix.  Therefore, we only have to check the entries of
the right hand term $\tilde{X}$. We add the trivial expression  $0 \,P_\B$ 
to $\tilde{X}$. The projections $P_\T$, $P_\S$, $P_1$, $P_\H$, and $P_\B$  
correspond  to a partition of the 
{\it basis states} of $(\cH_p\otimes \cH_d \otimes \cH_a)^{\otimes 3}$
into equivalence classes. The corresponding partition of the rows of $\tilde{X}$ decomposes $\tilde{X}$ into 
the submatrices $W_\H U_\T P_\T$,  $W_\H U_\S P_\S$, $W_\H P_1$,  $U_\H P_\H$ plus some trivial rows (corresponding to $P_\B$) 
that contain only zeros. 
The non-trivial submatrices, are in turn, submatrices of  the unitary matrices  $W_\H U_\T$, $W_\H U_\S$, $W_\H$, and $U_\H$.
Since they all contain exactly one Hadamard gate and only permutations apart from this, the  statement is clearly true.  

The corresponding Hamiltonian $\tilde{H}$ is 
\[
\tilde{H}:=\sum_j (\tilde{V}_j + \tilde{V}_j^\dagger)\,,
\]
where $\tilde{V}_j$ denote the copies of $\tilde{V}$ acting on cells $j-1$, $j$, and $j+1$. It is important that the entries of $\tilde{H}^m$ still indicate the result of the computation. Set
\[
|\tilde{\alpha}\>:=
|\alpha\> \otimes |0\>^{\otimes L} \quad \hbox{ and } \quad  
|\tilde{\omega}\> :=
|\omega\>\otimes |0\>^{\otimes L}\,,
\]
where the state $|0\>^{\otimes L}$ denotes the initial state of the auxiliary band. It is readily verified that
\[
\<\tilde{\alpha}|\tilde{H}^m|\tilde{\omega}\> = \frac{1}{\sqrt{1+d}} \<\alpha|H^m|\omega\> = \frac{1}{\sqrt{1+d}} (p_0-p_1) (\cP^m)_{0,\ell-1}\,,
\]
where the parameter $d$ is set to $0$ if the number of dummy Hadamard gates is even and to $1$ otherwise. In other words, an odd number of dummy Hadamard gates decreases the overlap by the factor $1/\sqrt{2}$.

%
%

We now convert $\tilde{V}$ to $\hat{V}$. The first idea to do that is as follows. Observe that the Pauli matrix $\sigma_x$ applied to $|-\>$ acts like a multiplication with $-1$. To use this observation for simulating negative signs, we introduce an additional band  $\cH_s^{\otimes L}$ where $\cH_s:=\C^2$.  We refer to $\cH_s$ as ``minus-sign-simulator'' or ``simulator'' for short. 

We write the logical Hadamard gate $U_\H$ in eq.~(\ref{tildeV}) in the form
\[
U_\H = {\bf 1}_p^{\otimes 3} \otimes \big(\sum_{ij} \mu_{ij} |i\>\<j|\big) \otimes {\bf  1}_a^{\otimes 3}\,,
\]
where
$i,j$ denote the basis states in $\cH_d^{\otimes 3}$ and $\mu_{ij}\in \{0,\pm 1/\sqrt{2}\}$. 
The symbols ${\bf 1}_p$ and ${\bf 1}_a$ denote the identity on program and ancilla cells, respectively.

We define the operators $M_{ij}$ acting on $\cH_s^{\otimes 3}$ by
\[
\begin{array}{rcll}
M_{ij} & = & I\otimes \sigma_x \otimes I & \mbox{if $\mu_{ij}=-1/\sqrt{2}$} \\
M_{ij} & = & I\otimes I \otimes I & \mbox{if $\mu_{ij}=1/\sqrt{2}$}  \\
M_{ij} & = & {\bf 0}              & \mbox{if $\mu_{ij}=0$}
\end{array}
\]
We now define the modified operator $\hat{U}_\H$ acting non-trivially on $\cH_d^{\otimes 3} \otimes \cH_s^{\otimes 3}$
(but formally  defined on  $\cH_p^{\otimes 3}\otimes \cH_d^{\otimes 3} \otimes \cH_a^{\otimes 3}\otimes \cH_s^{\otimes 3}$)
by
\[
\hat{U}_\H := {\bf 1}_p^{\otimes 3}\otimes \big( \sum_{ij} |i\>\<j| \otimes M_{ij} \big)\otimes {\bf 1}_a^{\otimes 3}\,.
\]
In the same way, we  modify the dummy Hadamard gate to $\hat{W}_\H$. 

If we defined a modified version $\hat{V}$ of $\tilde{V}$ by replacing $W_\H$ and $U_\H$ with $\tilde{W}_\H$ and $\tilde{U}_\H$, respectively (and replace the remaining operators in eq.~(\ref{tildeV}) with their canonical embeddings into the extended system) we would already obtain an operator with entries $0,1$. It would act just like $\tilde{V}$ provided that the simulator cell in the middle is in the state $|-\rangle$.  To 
simulate the Hamiltonian $\tilde{H}$  by a modified version $\hat{H}$ in this manner we would then have to initialize every simulator cell in the state $|-\rangle$. However, this would only allow us to reconstruct the original overlap $\<\tilde{\alpha}|\tilde{H}^m|\tilde{\omega}\>$ by extending $|\tilde{\alpha}\>$ and $|\tilde{\omega}\>$ with the multiple superposition $|-\>^{\otimes L}$.  This overlap would correspond to a linear combination of an exponentially large number of entries of $\hat{H}^m$ 
as opposed to our intention to reduce the problem to comparing only {\it two} entries. For  this reason, we initially prepare only {\it one} simulator cell in the superposition $|-\>$ and propagate the superposition to the ``active'' cell, i.e., the cell in which either of the Hadamard gates is currently executed.
In analogy to the swapping on the auxiliary band we define 
\[
\hat{V}:= 
C \,
\big(S_{0,1}S'_{0,1}\, Q_r + Q_s + S_{-1,0} S'_{-1,0} Q_l\big)
 \,  
\big(\hat{W}_\H (U_\T P_\T + U_\S P_\S + P_1) + \hat{U}_\H P_\H \big)\,,
\]
where the swaps $S'$ act on the simulator band. The particular form of $\hat{V}$ ensures that it 
only contains $0,1$ as entries. 
To show this formally 
we use similar  arguments as above and recall  that the term
\[
\hat{C}:=
C \, 
\big(S_{0,1}S'_{0,1}\, Q_r + Q_s + S_{-1,0} S'_{-1,0} Q_l\big)
\] 
is only a permutation matrix.
For the remaining term 
\[
\hat{X}:=\hat{W}_\H(U_\T P_\T+U_\S P_\S+P_1)+\hat{U}_\H P_\H
\]
we again define a decomposition into submatrices
$\hat{W}_\H U_\T P_\T$, $\hat{W}_\H U_\S P_\S$, $\hat{W}_\H P_1$, and  $\hat{U}_\H P_\H$.  They are, in turn, submatrices of
$\hat{W}_\H U_\T$, $\hat{W}_\H U_\S$, $\hat{W}_\H$, and  $\hat{U}_\H$. These matrices certainly have only 
the entries $0,1$ since  they have been obtained from
matrices with entries $0,\pm 1/\sqrt{2}$ by the rule described  above.  

Note that $\hat{V}+\hat{V}^\dagger$ is an adjacency matrix because there is no basis state $|b\rangle$ of
$\cH^{\otimes 3}_p \otimes \cH_d^{\otimes 3}\otimes \cH_a^{\otimes 3} \otimes \cH_s^{\otimes 3}$  
such that $\hat{V}|b\rangle \neq 0$ and $\hat{V}^\dagger|b\rangle \neq 0$. This  can easily be  checked by verifying
that there is no configuration in cell $-1$, $0$, and $1$ of the program and data bands such that one transition rule 
applies in forward {\it and } another (or the same) transition rule applies
in backward direction. 

The matrix $\hat{V}+\hat{V}^\dagger$ defines a relation on the substrings of length three over the alphabet $\cA$ of $224$ symbols that correspond to the basis states of $\cH_p\otimes \cH_d\otimes \cH_a\otimes \cH_s$ as follows.  Let $\sigma,\tau\in\cA^3$ be two arbitrary substrings.  Then $\sigma$ can be replaced by $\tau$ and vice versa iff $\<\sigma|\hat{V}+\hat{V}^\dagger|\tau\> = 1$. 

This collection of allowed replacements of substrings of length $3$ gives rise to an adjaceny matrix $A$ with strings in $\cA^L$ as vertices in the following way: $A_{s,t}=1$ iff there is a triple of consecutive positions $j-1,j,j+1$ with $j=2,\dots,L-1$ such that (1) $s$ and $t$ differ at most at these positions (2) the substring $s_{j-1} s_j s_{j+1}$ can be replaced by the substring $t_{j-1} t_j t_{j+1}$ in the sense of the relation specified by $\hat{V}+\hat{V}^\dagger$.  

The modified Hamiltonian $\hat{H}$ is 
\[
\hat{H}=\sum_j (\hat{V}_j +\hat{V}_j^\dagger)\,.
\]
Note that this Hamiltonian can have entries greater than $1$ in contrast to $A$. For example, this can occur if we can go from some string $s$ to some other string $t$ by applying two different substitution rules on two different triples. But it is clear that $H_{s,t}>0$ iff $A_{s,t}=1$.  However, the adjacency matrix $A$ and the Hamiltonian $\hat{H}$ are equal when restricted to the span of $F^i|\alpha\>$. This is because the transition rules and the initial state were defined such that there is always exactly one possible transition in forward time direction (see the properties of the states $F^i|\alpha\>$ that are discussed at the beginning of Subsection~\ref{subsec:encodingPowers}).

Appropriate entries of $A^m$  encode the solution of  the computation.
To see this  
we define the states 
\begin{eqnarray*}
|\alpha_0\> & := & |\tilde{\alpha}\> \otimes |0\cdots 000 \cdots 0\> \\
|\alpha_1\> & := & |\tilde{\alpha}\> \otimes |0\cdots 010 \cdots 0\> \\
|\omega_0\> & := & |\tilde{\omega}\> \otimes |0\cdots 000 \cdots 0\> \\
|\omega_1\> & := & |\tilde{\omega}\> \otimes |0\cdots 010 \cdots 0\>\,,
\end{eqnarray*}
where the tensor components on the right hand side refer to the simulator band and the symbol $1$ is above the execution symbol in $|\tilde{\alpha}\>$ and $|\tilde{\omega}\>$, respectively.  

Using the superpositions $|\alpha_\pm\>:=(|\alpha_1\> \pm |\alpha_0\>)/\sqrt{2}$ and $|\omega_\pm \> := |\omega_1\> \pm
|\omega_0\>)/\sqrt{2}$ we can reproduce the relevant entry of $\tilde{H}^m$:
\begin{equation}\label{overH}
\< \alpha_-|A^m |\omega_- \> = 
\< \alpha_-|\hat{H}^m|\omega_-\> = 
\sqrt{2}^m \,\< \tilde{\alpha}|\tilde{H}^m|\tilde{\omega}\>\,. 
\end{equation}
The factor $\sqrt{2}^m$ occurs because the entries $1/\sqrt{2}$ and $-1/\sqrt{2}$ in $\tilde{H}$ have been replaced by the matrices $I$ and $\sigma_x$, respectively, that have eigenvalues of modulus $1$.  Note that we have $\<\alpha_- |\hat{H}^m |\omega_+\>=0$ because 
the computation never transforms the superposition $|-\rangle$ into $|+\>$ in the simulator band. Using $|\omega_0\> = (|\omega_+\> + |\omega_-\>)/\sqrt{2}$ we obtain $\< \alpha_-|\hat{H}^m|\omega_-\> = \sqrt{2}\<\alpha_-|\hat{H}^m|\omega_0\>$.  Putting everything together, we obtain the desired result
\begin{eqnarray}
\<\alpha_0|A^m|\omega_0\> - \<\alpha_1|A^m|\omega_0\> & = &
\<\alpha_0|\hat{H}^m|\omega_0\> - \<\alpha_1|\hat{H}^m|\omega_0\> \\
& = &
\sqrt{2}^m \<\tilde{\alpha}|\tilde{H}^m|\tilde{\omega}\> \\
& = &
\sqrt{2}^m\, (p_0-p_1)\, \frac{1}{\sqrt{d+1}}\,(\cP^m)_{0,\ell-1}\,.
\end{eqnarray}

The  basis states $|\omega\rangle$, $|\alpha_0\rangle$, and $|\alpha_1\rangle$ define our strings of interest via 
$s:=\omega$, $t:=\alpha_1$, and  $t':=\alpha_0$. Then we have 
\begin{eqnarray}
\Delta_{s,t,t'}(m) & = & (A^m)_{st}-(A^m)_{st'} \nonumber \\
& = &
\langle \alpha_0| A^m|\omega_0\rangle - \langle \alpha_1|A^m|\omega_0\rangle \nonumber\\ 
&=& 
\sqrt{2}^m\, (p_0-p_1)\, \frac{1}{\sqrt{d+1}}\,(\cP^m_\ell)_{0,\ell-l}  \label{DeltaAlsPm}\,,
\end{eqnarray}
i.e., the sign of $\Delta_{s,t,t'}(m)$ depends on the solution computed by the original circuit. We have to show that the gap and growth conditions are satisfied to complete the proof that the so constructed instance of our string rewriting problem determines whether a given string $x\in\Pi$ is either in $\Pi_{{\rm YES}}$ or in $\Pi_{{\rm NO}}$.

\subsection{Appropriate choice of the power $m$}

To choose $m$ appropriately and to check that both conditions are satisfied, we use the spectral decomposition of the adjacency matrix of the line graph
\[
\cP = \sum_{j=0}^{\ell-1} \lambda_j |e_j\> \<e_j|\,.
\]
The eigenvalues are $\lambda_j=2\cos\big(\frac{\pi (j+1)}{\ell + 1}\big)$ for $j=0,\ldots,\ell-1$.  The normalized eigenvecotrs are 
\[
|e_j\> := \sum_{k=0}^{\ell -1} e_{j,k} |k\>
\]
with
\[
e_{j,k} := \sqrt{\frac{2}{\ell+1}} \sin \Big( \frac{\pi(j+1)(k+1)}{\ell+1} \Big) 
\]
for $j,k=0,\ldots,\ell-1$ \cite{Cvetkovic}.

Define the (signed) weights 
\[
w_j:=\<0|e_j\>\<e_j|\ell-1\>= e_{j,0} \, e_{j,\ell-1} \quad \hbox{ for } j=0,\ldots,\ell-1\,.
\]
We have 
\[
(\cP^m)_{0,\ell-1} = \sum_{j=0}^{\ell-1} \lambda_j^m \, w_j\,.
\]
The upper bound
\begin{equation}\label{eq:upperBound}
(\cP^m)_{0,\ell-1} \le \lambda_0^m
\end{equation}
follows from $|\lambda_j|\le \lambda_0$ for $j=0,\ldots,\ell-1$ and 
\[
\sum_{j=0}^{\ell-1} | w_j | \le 1\,.
\]
The latter bound follows by elementary linear-algebraic arguments from the fact that the eigenvectors $|e_j\>$ are orthonormal.

We obtain the growth condition
\[
|\Delta_{s,s',t}(n)| \leq \frac{1}{\sqrt{d+1}} \big( \sqrt{2}\cdot \lambda_1 \big)^n\,.
\]
by applying the upper bound on $(\cP^m)_{0,\ell-1}$ in eq.~(\ref{eq:upperBound}) to eq.~(\ref{DeltaAlsPm}).

To derive a lower bound on $(\cP^m)_{0,\ell-1}$, we consider the two eigenvalues of largest modulus, i.e., $\lambda_0=2\cos(\pi/\ell +1)$ and $\lambda_{\ell-1}=-2\cos (\pi/\ell+1)$. The corresponding weights $w_0$ and $w_{\ell-1}$ satisfy the relation
\[
w_{\ell-1} = (-1)^{\ell-1} w_0
\]
since 
\[
\sin \Big(\frac{\pi\ell^2}{\ell+1}\Big) = 
\sin \Big((\ell-1)\pi + \frac{\pi}{\ell+1}\Big) =
(-1)^{\ell-1} \sin\Big(\frac{\pi}{\ell+1}\Big)\,.
\]
Recall that $m$ is even if $\ell$ is odd and vice versa so that $\lambda_0 w_0 = \lambda_{\ell-1} w_{\ell-1}>0$.  Observe that $|\lambda_j|\le \lambda_1$ for $j=1,\ldots,\ell-2$.

We now obtain the lower bound
\begin{eqnarray}
(\cP^m)_{0,\ell-1} 
& = & 
2 \lambda_0^m w_0 + \sum_{j=1}^{\ell-2} \lambda^m_j w_j \nonumber \\
& \ge & 
2 \lambda_0^m w_0 - \lambda_1^m \sum_{j=1}^{\ell-2} |w_j| \nonumber  \\
& \ge &
2 \lambda_0^m w_0 - \lambda_1^m  \nonumber \\
& = &
\lambda_0^m \left( 2 w_0 - \Big(\frac{\lambda_1}{\lambda_0}\Big)^m \right) \nonumber \\
& \ge &
\lambda_0^m w_0 \label{w0}
\end{eqnarray}
if $m$ is chosen such that $(\lambda_1/\lambda_0)^m \leq w_0$. We set $\tilde{\ell}:=\ell+1$  
and $m:=\tilde{\ell}^3$ and obtain
\begin{eqnarray*}
\left(\frac{\lambda_1}{\lambda_0}\right)^m  &\leq& \left( \frac{\cos (2 \pi/\tilde{\ell})}{\cos (\pi/\tilde{\ell})} \right)^m \\
                                            &\leq& \left( \frac{1-(\pi/\tilde{\ell})^2}{1-\frac{1}{2}(\pi/\tilde{\ell})^2} \right)^m \\
                                            &=& \left[
         \left( \frac{1-(\pi/\tilde{\ell})^2}{1-\frac{1}{2}(\pi/\tilde{\ell})^2} \right)^{\tilde{\ell}^2} \right]^{\tilde{\ell}} \\
\end{eqnarray*}
The term inside the square brackets converges to $e^{-\pi^2/2}$, hence the entire term converges  to zero.
Inserting eq.~(\ref{w0}) into eq.~(\ref{DeltaAlsPm})
 we obtain
\[
|\Delta_{s,s',t}(m)| \ge \frac{1}{3} w_0\, \frac{1}{\sqrt{d+1}}(\sqrt{2} \cdot \lambda_1)^m\,,
\]
observing
$|p_0-p_1|\geq 1/3$
due to the
promise that either $p_0\geq 2/3$ or $p_0\leq 1/3$.
Using the Taylor expansion of the sine function one verifies easily that $w_0$ converges only inverse polynomially to zero.
With $\epsilon:=w_0$  
the gap  condition in the sense of Definition~\ref{def:stringRewriting} is satisfied.

\section{Conclusions}

We have described a purely classical combinatorial problem that characterizes the complexity class PromiseBQP, i.e., the class
of problems that can be solved efficiently on a quantum computer. Given that BQP$\neq$ BPP, our result shows that the quantum computer outperforms the classical computer in estimating differences of combinatorial
quantities.  

\section*{Acknowledgments}
We would like to thank Daniel Nagaj for very helpful discussions about the transition rules.  P.W. gratefully acknowledges the support by the NSF grant CCF-072677.


\begin{thebibliography}{10}

\bibitem{Goldreich}
O.~Goldreich.
\newblock On promise problems.
\newblock {\em Electronic Colloquium on Computational Complexity}, (18), 2005. \\
\newblock {\tt http://eccc.hpi-web.de/eccc-reports/2005/TR05-018/index.html}


\bibitem{DiagonalEntryToC}
D.~Janzing and P.~Wocjan.
\newblock {A simple PromiseBQP-complete matrix problem}.
\newblock {\em Theory of Computing}, 3:61--79, 2007.

\bibitem{KnillQuadr}
E.~Knill and R.~Laflamme.
\newblock Quantum computation and quadratically signed weight enumerators.
\newblock {\em Inf. Process. Lett.}, 79(4), 2001.

\bibitem{PawelYard}
P.~Wocjan and J.~Yard.
\newblock The {Jones} polynomial: quantum algorithms and applications in
  quantum complexity theory.
\newblock {\em Quantum Information \& Computation}, 
\newblock Vol. 8, No. 3 \& 4, pp. 147--180, 2008.


\bibitem{aharonov-2006-}
D.~Aharonov and I.~Arad.
\newblock The {BQP}-hardness of approximating the {Jones} polynomial.
\newblock quant-ph/0605181, 2006.

\bibitem{Freedman}
M.~Freedman, A.~Kitaev, and Z.~Wang.
\newblock Simulation of topological field theories by quantum computers.
\newblock {\em Comm. Math. Phys.}, 227(3):587--603, 2002.

\bibitem{Jones}
D.~Aharonov, V.~Jones, and Z.~Landau.
\newblock A polynomial quantum algorithm for approximating the {Jones}
  polynomial.
\newblock 
Proceedings of the 38th Annual ACM Symposium on Theory of Computing, Seattle 2006.
pp 427-436.


\bibitem{AharonovTutte}
D.~Aharonov.
\newblock {Polynomial quantum algorithms for additive approximations of the
  Potts model and other points of the Tutte plane}.
\newblock quant-ph/0702008.

\bibitem{FR98}
L.~Fortnow and J. Rogers. 
Complexity limitations on quantum computation.
{\em Proc. of IEEE Complexity'98},
pp. 202-209, 1999.


\bibitem{NumberOfWalks}
D.~Janzing, P.~Wocjan.
\newblock
 BQP-complete Problems Concerning Mixing Properties of Classical Random Walks on Sparse Graphs.
\newblock quant-ph/0610235.


\bibitem{ATS}
D.~Aharonov and A.~Ta-Shma.
\newblock Adiabatic quantum state generation and statistical zero knowledge.
\newblock In {\em Proc. 35th Annual ACM Symp. on Theory of Computing}, pages
  20--29, 2003.

\bibitem{ChildsDiss}
A.~Childs.
\newblock {\em Quantum information processing in continuous time}.
\newblock PhD thesis, Massachusetts Institute of Technology, 2004.

\bibitem{BACS:06}
{D. W.} Berry, G.~Ahokas, R.~Cleve, and {B. C.} Sanders.
\newblock Efficient quantum algorithms for simulating sparse {Hamiltonians}.
\newblock {\em Comm. Math. Phys.}, 270(2):359--371, 2007.


\bibitem{Vollbrecht}
K.~Vollbrecht and I.~Cirac.
\newblock Quantum simulators, continuous-time automata, and translationally
  invariant systems.
\newblock quant-ph/0704.343.

\bibitem{Benioff}
P.~Benioff.
\newblock {The computer as a physical system: A microscopic quantum mechanical
  model of computers as represented by Turing machines}.
\newblock {\em J. Stat. Phys.}, 22(5):562--591, 1980.

\bibitem{Feynman:85}
R.~Feynman.
\newblock Quantum mechanical computers.
\newblock {\em Opt. News}, 11:11--46, 1985.

\bibitem{Margolus:90}
N.~Margolus.
\newblock Parallel quantum computation.
\newblock In W.~Zurek, editor, {\em Complexity, Entropy, and the Physics of
  Information}. Addison Wesley Longman, 1990.

\bibitem{Ergodic}
D.~Janzing and P.~Wocjan.
\newblock Ergodic quantum computing.
\newblock {\em Quant. Inf. Process.}, 4(2):129--158, 2005.

\bibitem{ErgodicQutrits}
D.~Janzing.
\newblock {Spin-1/2 particles moving on a 2D lattice with nearest-neighbor
  interactions can realize an autonomous quantum computer}.
\newblock {\em Phys. Rev. A} 75:012307, 2007.

\bibitem{KitaevShen}
A.~Kitaev, A.~Shen, and M.~Vyalyi.
\newblock {\em Classical and Quantum Computation}, volume~47.
\newblock Am. Math. Soc., Providence, Rhode Island, 2002.

\bibitem{Kempe2local}
J.~Kempe, A.~Kitaev, and O.~Regev.
\newblock {The complexity of the local Hamiltonian problem}.
\newblock {\em Proc. 24th FSTTCS, accepted to SICOMP}, 2004.

\bibitem{Oliveira}
R.~Oliveira and B.~Terhal.
\newblock The complexity of quantum spin systems on a two-dimensional square
  lattice.
\newblock  quant-ph/0504050v.

\bibitem{GottesmanQMA}
D.~Gottesman, D.~Aharonov, S.~Irani, and J.~Kempe.
\newblock 
The power of quantum systems on a line. FOCS'2007.


\bibitem{CMP} D.~Janzing, P.~Wocjan, and S.~Zhang.
\newblock 
A single-shot measurement of the energy of product states in a translation invariant spin chain can replace any quantum computation.
\newblock arXiv:0710.1615.


\bibitem{Feynman}
R.~Feynman.
\newblock {\em Feynman Lectures on Computation}.
\newblock Perseus Pr., 1996.

\bibitem{aharonov-2003}
D.~Aharonov.
\newblock A simple proof that {Toffoli} and {Hadamard} are quantum universal.
\newblock quant-ph/0301040, 2003.

\bibitem{Cvetkovic}
D.~Cvetkovic.
\newblock {\em Eigenspaces of graphs}.
\newblock Cambridge University Press, 1997.

\end{thebibliography}

\end{document}